\documentclass[eqsecnum, aps, pra, amsmath, amssymb, onecolumn, superscriptaddress, 11pt ]{revtex4-1}

\usepackage{graphicx}
\usepackage{dcolumn}
\usepackage[colorlinks=true,
            linkcolor=blue,
            urlcolor=black,
            citecolor=blue]{hyperref}
\usepackage{bm}
\usepackage{enumerate}
\usepackage{lipsum}
\usepackage{threeparttable}
\usepackage{epstopdf}
\usepackage{float}
\usepackage{xcolor,colortbl}
\usepackage[caption = false]{subfig}
\usepackage{tabularx}
\usepackage{chngpage}

\begin{document}


\title{Vector polarizability of  atomic state induced by a linearly polarized vortex beam:  External control of  magic, tune-out wavelengths, and heteronuclear spin oscillations}
\author{Anal Bhowmik}
\email{abhowmik@campus.haifa.ac.il}
\affiliation{Haifa Research Center for Theoretical Physics and Astrophysics, University of Haifa, Haifa 3498838, Israel}
\affiliation{Department of Mathematics, University of Haifa, Haifa 3498838, Israel}
\author{Narendra Nath Dutta}
\affiliation{Department of Chemical Sciences, Indian Institute of Science Education and Research Mohali, Punjab-140306, India.}
\author{Sonjoy Majumder}
\email{sonjoym@phy.iitkgp.ernet.in}
 \affiliation{Department of Physics, Indian Institute of Technology Kharagpur, Kharagpur-721302, India.}






\begin{abstract}
Experiments with vortex beams have shown a surge of interest in controlling cold atoms.  Most of the controlling protocols are dominated by circularly polarized light due to its ability to induce vector polarization at atoms, which is impossible for paraxial linearly polarized light. Here we develop a theory for frequency dependent polarizability of an atomic state interacting with a focused linearly polarized vortex beam. The naturally induced spin-orbit coupling in this type of linearly polarized beam produces vector component of the valence polarizability to an atomic state,  obeying the total angular momentum conservation of the beam.   The theory is employed on $^{87}$Sr$^{+}$ ion to accurately calculate the magic wavelengths for the clock transitions  and tune-out wavelengths for the clock states using  relativistic coupled-cluster method. The induced vector component in the dynamic polarizability due to the linearly polarized focused vortex beam promotes fictitious magnetic field to the atomic state.  We demonstrate that this fictitious magnetic field, depending on the focusing angle and OAM of the beam, improves the flexibility of the coherent heteronuclear spin oscillations in a spin-1 mixture of  $^{87}$Rb and $^{23}$Na atoms.
\end{abstract}

    \vspace{-3.5cm}                          
\maketitle

\section{INTRODUCTION}
In  recent years, laser trapping and cooling of neutral atoms  have  attracted significant attention to the experimentalists and achieved as an established technique in high precision spectroscopic measurements \cite{Champenois2004, Chou2010, Ludlow2015}. Along with neutral atoms, optical dipole trappings of ions have recently surged interest in ultra-cold  physics \cite{Laupretre2019,Schmidt2018,Lambrecht2017,Schaetz2017}. However, mechanism of the trapping by optical means inevitably produces the Stark shifts in the atomic energy levels and  influences the fidelity of the precision measurements. Magic wavelengths are the unique wavelengths of the external laser beam for which the differential ac-Stark shift of an atomic transition effectively vanishes. Therefore, the impediment in precise spectroscopic measurements can be eliminated almost entirely if the atoms are confined at the predetermined magic wavelengths of the laser beam. The light at magic wavelengths have significant applications in atom optics, such as atomic interferometers \cite{Biedermann2015}, atomic clocks \cite{Margolis2009, Rosenbusch2009, Nicholson2015},  and atomic magnetometers \cite{Dong2015}, etc.

Determinations  of magic wavelengths for an atomic transition depend mainly on how accurately the  frequency-dependent  valence polarizability (POL) values of the atomic Zeeman sub-levels are calculated. The valence POL consists of three components, scalar POL ($\alpha_V^0$), vector POL ($\alpha_V^1$), and tensor POL ($\alpha_V^2$)  \cite{Flambaum2008, Arora2007, Lacroute2012, Arora2012}.  In general, POL is defined in terms of an off-resonant electric dipole interaction between the atom and its trapping light. The vector POL, which yields the energy shift among the Zeeman sublevels, arises usually due to  a circularly polarized light \cite{Flambaum2008, Bhowmik2018, Das2020}. It is one of the non-magnetic field sources to remove Zeeman degeneracy  and becomes  very significant in the evaluation of magic wavelengths for cases such as "knob" to adjust an optical trap \cite{Kim2013}.

In this work, we theoretically develop and demonstrate that a linearly polarized focused vortex (LP-FV) beam can naturally produce  vector part of the  dynamic POL when it interacts with cold atoms or ions. The vector part is certainly not a component of dynamic POL for a linearly polarized Gaussian or paraxial vortex beam. The distinctive feature of the vortex beam is that in addition to the spin angular momentum  associated with the polarization, it has orbital angular momentum (OAM) which arises due to the helical phase front of the beam \cite{Allen1992, Babiker2002, Lembessis2011, Afanasev2013, Mondal2014, Bhowmik2016, Schmiegelow2016,  Bhowmik_JPC_2018, Romero2012, SDas2020, Quinteiro2020}.  We find that the origin of  the vector POL comes from the prolific artifacts of the spin-orbit coupling of LP-FV beam satisfying the conservation of  the total angular momentum of the beam during the interaction-process.   Also, this vector POL creates  a fictitious magnetic field \cite{Kien2013} when it interacts with an atomic system. This fictitious magnetic field interrogates, manipulates, and traps an atom or ion independently \cite{Albrecht2016} or in conjunction with the real magnetic field \cite{Schneeweiss2014}. Further, this fabricated field provides a unique opportunity for exploring spin-dynamics in the form of coherent spin oscillations \cite{Stenger1998, Chang2004, Chang2005, Klempt2009} in the ultra-cold spinor quantum gas.  This has been the subject of intensive theoretical and experimental research due to its high controllability \cite{Kurn2013}.

The primary objective of the present work is to visualize the effect of vector POL, originated due to the spin-orbit coupling of the LP-FV beam, in the clock  transitions and clock states.    We apply our theory to calculate the dynamic POL values of the fine-structure states, $5S_{\frac{1}{2}}$, $4D_{\frac{3}{2}}$, and $4D_{\frac{5}{2}}$  of  $^{87}$Sr$^{+}$. Singly ionized strontium is  an excellent candidate for optical frequency standard, quantum  information, and quantum storage \cite{Akerman2015, Shapira2018}. As the spin-orbit coupling of LP-FV beam varies with the choice of  OAM and focusing angle of the lens, we quantitatively show the effects of these parameters on the  magic wavelengths for $5S_{\frac{1}{2}}\rightarrow 4D_{\frac{3}{2}, \frac{5}{2}}$ transitions and tune-out wavelengths (where the dynamic POL value becomes zero \cite{LeBlanc2007, Wang2016}). These quantifications will be important for the experimentalists to choose the parameters of the vortex beams used for trapping \cite{Kim2013, Albrecht2016}. In order to manifest the impact of the vector POL, we compare the calculated magic and tune-out wavelengths induced from the LP-FV beam with the corresponding wavelengths induced from a linearly polarized Gaussian beam. Moreover, as an important application,  we  investigate how the fictitious magnetic field, originated from the  vector POl, can control the coherent heteronuclear spin oscillations in a spin-1 mixture of  $^{87}$Rb and $^{23}$Na atoms with influences from both linear and quadratic Zeeman shifts.  

\section{THEORY}
According to the time-independent second-order perturbation theory \cite{Mitroy2010},  the ac-Stark shift of an atomic state in an external oscillating electric field $\mathcal{E}(\omega)$ is expressed by $ \Delta F(\omega)= -\frac{1}{2}\alpha^T(\omega)\mathcal{E}^2$, where $\alpha ^T(\omega)$ is the total dynamic POL of the atomic  state at frequency $ \omega $ and $\mathcal{E}$ is the magnitude of the external electric field.  For a single-valence atom with a valence electron in the $v$th orbital, the total dynamic POL can be written as $\alpha^T(\omega)=\alpha^C(\omega)+\alpha^{VC}(\omega)+\alpha^V(\omega)$. Here $\alpha^C(\omega)$  is the  frequency-dependent  core POL of the ionic core  obtained by removing the valence electron. $ \alpha^{VC}(\omega) $ gives a  correction  \cite{Dutta2015} to the core POL due to the presence of the valence electron,  and  it is  considered as    $\omega$-independent in the present work.  $\alpha^V(\omega)$ is the valence POL of the single-valence state.    $\alpha^C$ and $\alpha^{VC}$ provide small contributions to $\alpha^T$ compared to $\alpha^V$, and they are  computed approximately using lower-order many-body perturbation theory  \cite{Mitroy2010, Ghosh1993, Dutta2020}.

 Now, we evaluate $\alpha^V(\omega)$ in the presence of an external LP-FV beam. As Laguerre-Gaussian (LG) beam is a well-established example of a vortex beam,  initially, we assume a  paraxial form of  linearly polarized coherent LG beam  without any off-axis node and propagates along the $z$-axis. The field  is expressed as  \cite{Mondal2014} $\boldsymbol{{{\mathcal{E}_i}}}(\rho, \phi, z, t)=\mathcal{E}_i(t) \left({{\sqrt{2}\rho }/{w_0 }}\right)^{\lvert l \rvert}e^{i(l\phi+k_0z)}
\boldsymbol{\hat{\textbf{x}}}$. Here $k_0$ is the wave number of the free space,  $w_0$ is the  waist-size  and $l$ is  OAM  of the beam. We consider that this paraxial LG beam is focused  by passing  through an objective (lens) with a high numerical aperture (NA) \cite{Bhowmik2016}. Then this focused LG beam interacts with a cold atom or ion  whose de Broglie wavelength is large enough to experience the intensity variation of this beam. In order to take a full advantage of the high NA of the lens, we assume that  $w_0$ overfills  the entrance aperture radius. According to the Kirchhoff's approximation in  diffraction theory \cite{Richards1959, Boivin1965}, the consequent components of the spin-orbit coupled LP-FV beam can be expressed as
\vspace{-0.3cm}
\begin{eqnarray}\label{eq1}
&\mathcal{E}&(\rho, \phi, z, t)= \left [\begin{array}{c} \mathcal{E}_x \\ \mathcal{E}_y\\ \mathcal{E}_z  \end{array}\right ] = (-i)^{l+1} \mathcal{E}_0 e^{i(l\phi-\omega t)}\nonumber \\&\times&\left [\begin{array}{c} u_l(\rho , z)+u_{l+2}(\rho , z)  e^{2i\phi}+u_{l-2}(\rho , z)  e^{-2i\phi}\\ -i(u_{l+2}(\rho , z)  e^{2i\phi}-u_{l-2}(\rho , z)  e^{-2i\phi})\\ -i(u_{l+1}(\rho , z)  e^{i\phi}-u_{l-1} (\rho , z) e^{-i\phi})  \end{array}\right ].
\end{eqnarray}
Here $\mathcal{E}_x$, $\mathcal{E}_y$, and $\mathcal{E}_z$ are  the $x$-, $y$-, and $z$- components of the electric field, respectively.  The amplitude of the focused electric field is $\mathcal{E}_0=\dfrac{\pi f}{\lambda} T \mathcal{E}_i$, where $\mathcal{E}_i$ is the amplitude of the incident electric field,  $T$ is the  transmission coefficient  of the objective,  and $f$ is its focal length related to $\rho$ by $\rho=f \sin\theta$ (Abbe sine condition). The coefficients $u_{l+m}$, where $m$ takes the values 0, $\pm1$, and $\pm2$   in the above expression, depend on the focusing angle of NA ($\vartheta_{m}$) following the relation \cite{Zhao2007} $u_{l+m}(\rho , z)=\int_0^{\vartheta_{m}}d\vartheta\left({\dfrac{\sqrt{2}\rho }{w_0 }}\right)^{\lvert l \rvert} \sin\vartheta 
\sqrt{\cos\vartheta}g_{\lvert m \rvert}(\vartheta) J_{l+m}(k\rho \sin\vartheta)e^{ikz\cos\vartheta}.$ Here $J_{l+m}(k\rho \sin\vartheta)$ represents cylindrical Bessel function and  $k=\mu k_0$ with $\mu$ as the refractive index of the lens medium.    $g_{|m|}(\vartheta)$ are the angular functions with $g_0 (\vartheta)=1+\cos\vartheta$, $g_1 (\vartheta)=\sin\vartheta$, and $g_2 (\vartheta)=1-\cos\vartheta$. 

Now, let us discuss about Eq.~(\ref{eq1}) in detail. Linearly polarized light can be considered as the superposition of left ($\beta=+1$) and right ($\beta=-1$) circularly polarized beams.  Because of the focusing and the diffraction from the edges of the aperture, each circularly polarized light ($\beta=\pm 1$) can be decomposed into three sets of local polarizations ($\pm 1$, $\mp 1$, and polarization along z-axis) \cite{Bhowmik2016}. Among these three sets of local polarizations, the first set has equal amplitude ($u_l$) for $\pm 1$ local polarizations. Therefore,  after passing through the focusing lens, the superposition of these local polarizations results linearly polarized beam with the OAM similar to the OAM of the incident beam. However, in the case of second set, different field amplitudes,  $u_{l+2}$ and $u_{l-2}$, are generated  with two different local polarizations ($-1$ and $+1$, respectively) and topological charges ($l+2$ and $l-2$, respectively, to conserve total angular momentum) after passing through the lens. Therefore, the field  gains  two opposite circular polarizations with different amplitudes and creates the vector part of valence POL in interaction with an atom or ion. The third set yields $u_{l+1}$ and $u_{l-1}$  fields with topological charges, $l+1$ and $l-1$, respectively. However,  both these fields are polarized along the $z$- direction, which is another interesting manifestation of focusing  beam.   Nevertheless,  using LP-FV beam  presented in Eq.~(\ref{eq1}),  $\alpha^V(\omega) $  of the atomic state can be presented by 
\begin{equation}\label{eq2}
\alpha^V(\omega) =C_0 \alpha_V^0(\omega)+ C_1\alpha_V^1(\omega)  + C_2\alpha_V^2(\omega).
\end{equation}
Here the coefficients $C_i$s are expressed in the following forms:
 \begin{eqnarray}\label{eq3}
  C_0& =& \{u_l\}^2+\{u_{l+1}\}^2+\{u_{l-1}\}^2+2[\{u_{l+2}\}^2+\{u_{l-2}\}^2],\nonumber \\ 
C_1&=& \left[2\{u_{l-2}\}^2-2\{u_{l+2}\}^2\right]\times\left(\frac{M_{J_V}}{2J_V}\right), \text{and} \\
C_2 &=& \left[\{u_l\}^2-\{u_{l+1}\}^2-\{u_{l-1}\}^2+2\{u_{l+2}\}^2 +  2\{u_{l-2}\}^2\right] \nonumber \\  &\times& \left({(3M_{J_V}^2-J_V(J_V+1))}/{(2J_V(2J_V-1))}\right),
\nonumber
\end{eqnarray}
 where $J_V$ and $M_{J_V}$ are the total angular momentum and its magnetic component for the single-valence atomic state $|\Phi_V\rangle$. The mathematical expressions of the scalar, vector,  tensor, and consequently the valence POL at the fine- and hyperfine-structure levels are presented in the Appendix. The calculation of $ \alpha^V(\omega)$  \cite{Mitroy2010, Bhowmik2018} of an atomic state directly  depends on different combinations of the integrals $u_{l+m}$. These integrals  can be altered with the various choices of the topological charges of the incident LG beam and the NA of the objective. As a consequence, the total POL values of the atomic states and the  magic wavelengths for the transitions among them can be tuned externally using different parameters  of the beam.

 \section{RESULTS AND DISCUSSIONS}

\begin{figure}[htb]
{\includegraphics[ scale=.30]{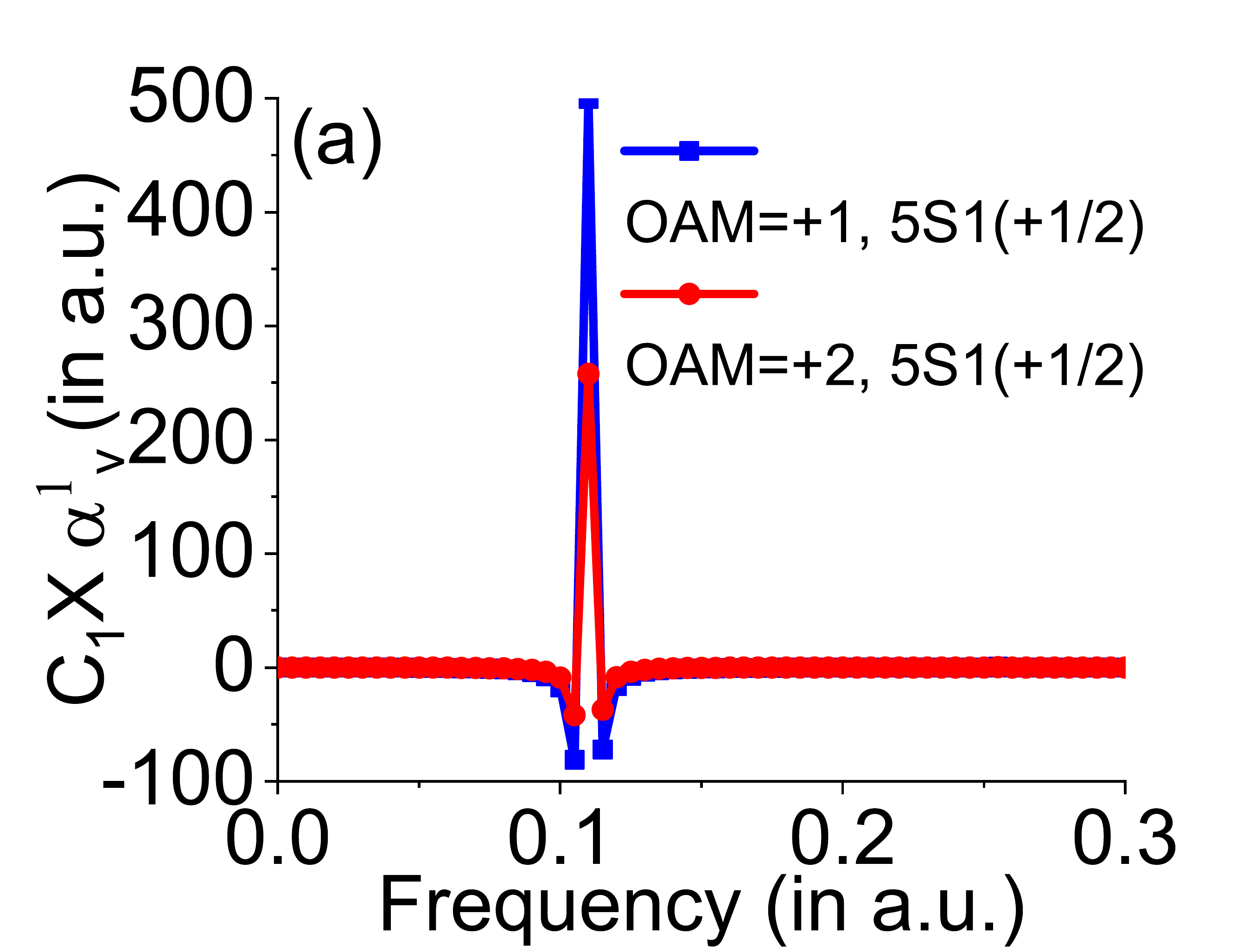}}
{\includegraphics[scale=.30]{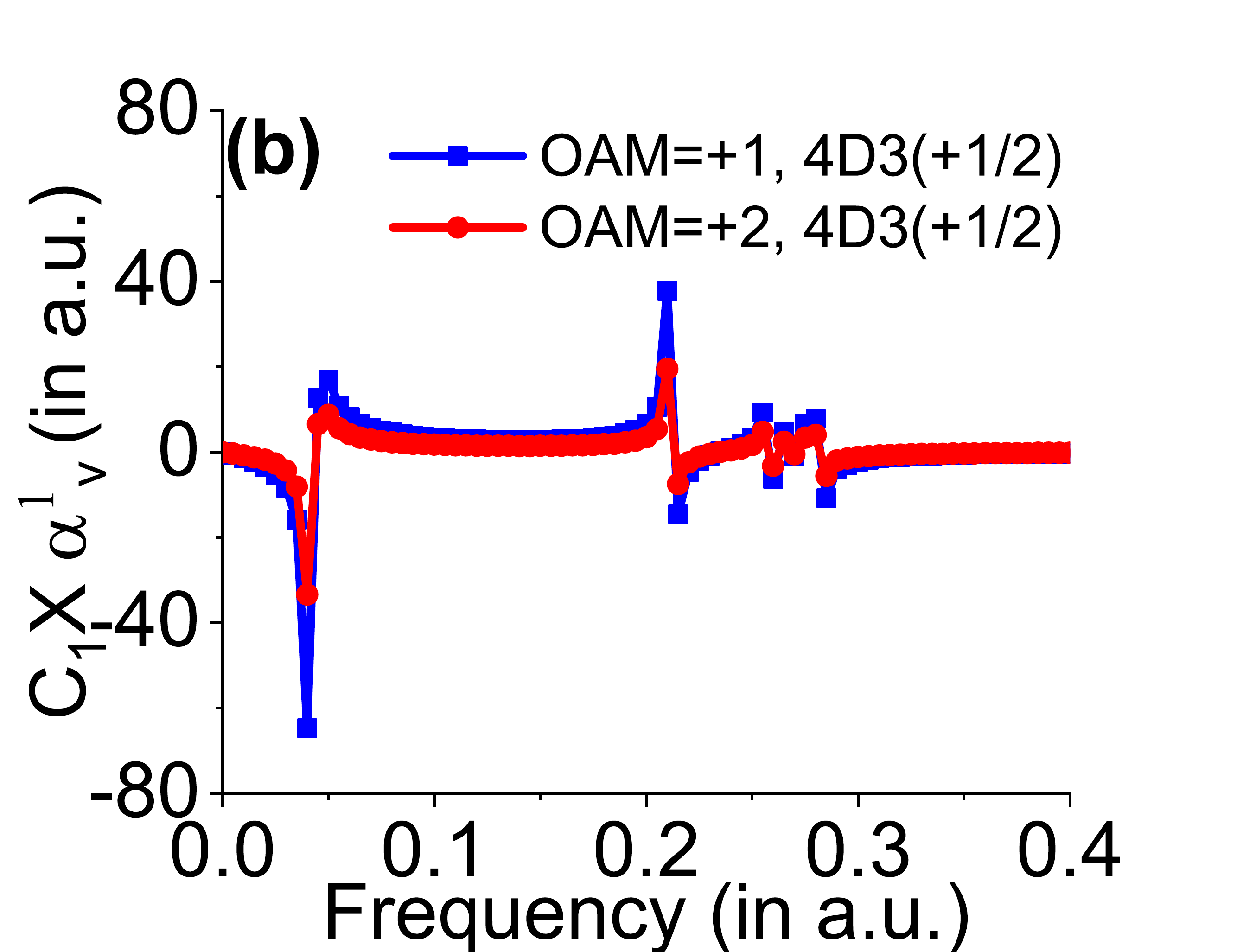}}
{\includegraphics[scale=.30]{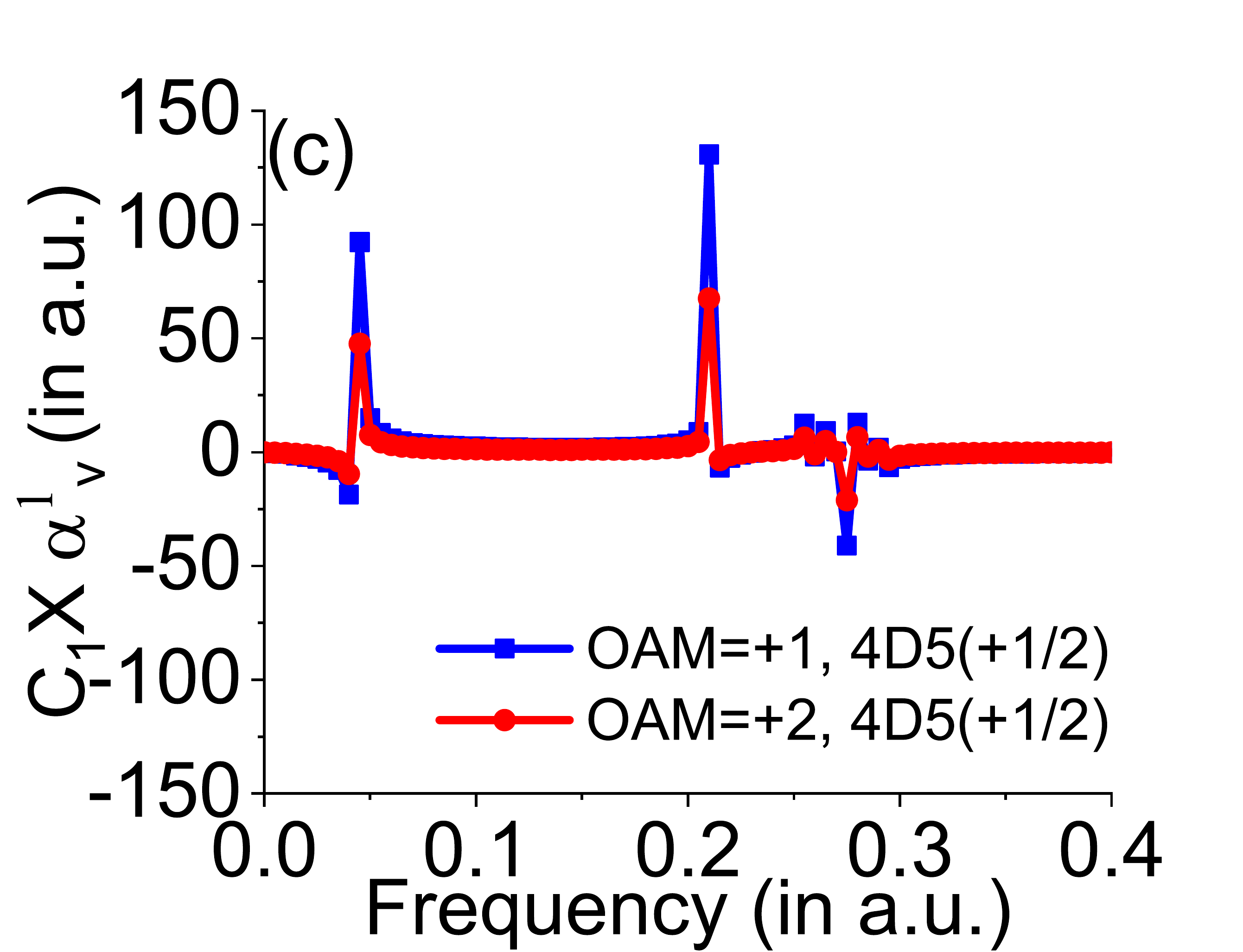}}
\caption{\label{fig1} Frequency  dependent  total vector POL ($C_1\alpha_V^1$)  of  $5S_{\frac{1}{2}}(+1/2)$ and $4D_{\frac{3}{2},\frac{5}{2}}(+ 1/2) $ states for LP-FV beam  with OAM=+1 and +2 at focusing angle 60$^\circ$.}
\end{figure}

As the electron-correlation most significantly affects  $\alpha^V(\omega)$ part of the dynamic POL  due to the loosely bound  valence electron, the accurate estimations of the scalar, vector, and tensor components of $\alpha^V(\omega)$ for the different states require correlation exhaustive many-body calculations \cite{Lindgren1985, Bishop1987, Lindgren1986, Lindgren_1985, Bartlett2007, Dutta2016,  Bhowmik2017, Bhowmik2017b,  Das2018}  with a sophisticated numerical approach (see Appendix for the details of computations).  To have a quantitative analysis of the effect of OAM of LP-FV beam on the total vector POL ($C_1\alpha_V^1(\omega)$) of an atomic state, we show variations of $C_1\alpha_V^1(\omega)$ with frequency for $5S_{\frac{1}{2}}(+1/2)$ and $4D_{\frac{3}{2},\frac{5}{2}}(+ 1/2) $ states of $^{87}$Sr$^{+}$  in Fig.~\ref{fig1} ($M_J=-\frac{1}{2}$ provides opposite sign to $C_1\alpha_V^1(\omega)$). The fine-structure states $5S_{\frac{1}{2}}$, $4D_{\frac{3}{2}}$, and $4D_{\frac{5}{2}}$ are indicated by $5S1$, $4D3$, and $4D5$, respectively in this figure and the next two figures of this paper. Here LP-FV beam has either OAM=+1 or +2 with focusing angle of 60$^\circ$.  Although the effect of variation of focusing angle to  $C_1\alpha_V^1(\omega)$  is marginal, it is significant to the total scalar and tensor POL values, and consequently to the total POL values.   The peak values of  $C_1\alpha_V^1(\omega)$  occur at  resonance frequencies: 0.11 a.u. for $5S_{\frac{1}{2}}$ state;  around 0.045 a.u and 0.21 a.u. for $4D_{\frac{3}{2}, \frac{5}{2}}$ states, respectively.  However, many small scale structures in $C_1\alpha_V^1(\omega)$ profiles of $4D_{\frac{3}{2},\frac{5}{2}}$ states appear due to multiple number of resonance transitions at around 0.25 a.u. to 0.30 a.u.

 It can be seen from the graphs that OAM=+1 always produces higher peak value of $C_1\alpha_V^1(\omega)$  compared to OAM=+2.  This is because of the stronger spin-orbit coupling for OAM=+1 compared to OAM=+2. Also, the magnitude  of the peak value  of $C_1\alpha_V^1(\omega)$  is higher for larger $M_J$ for a fixed value of $J$, which is obvious from the expression of $C_1$.    
\begin{figure}[htb]
	{\includegraphics[scale=.30]{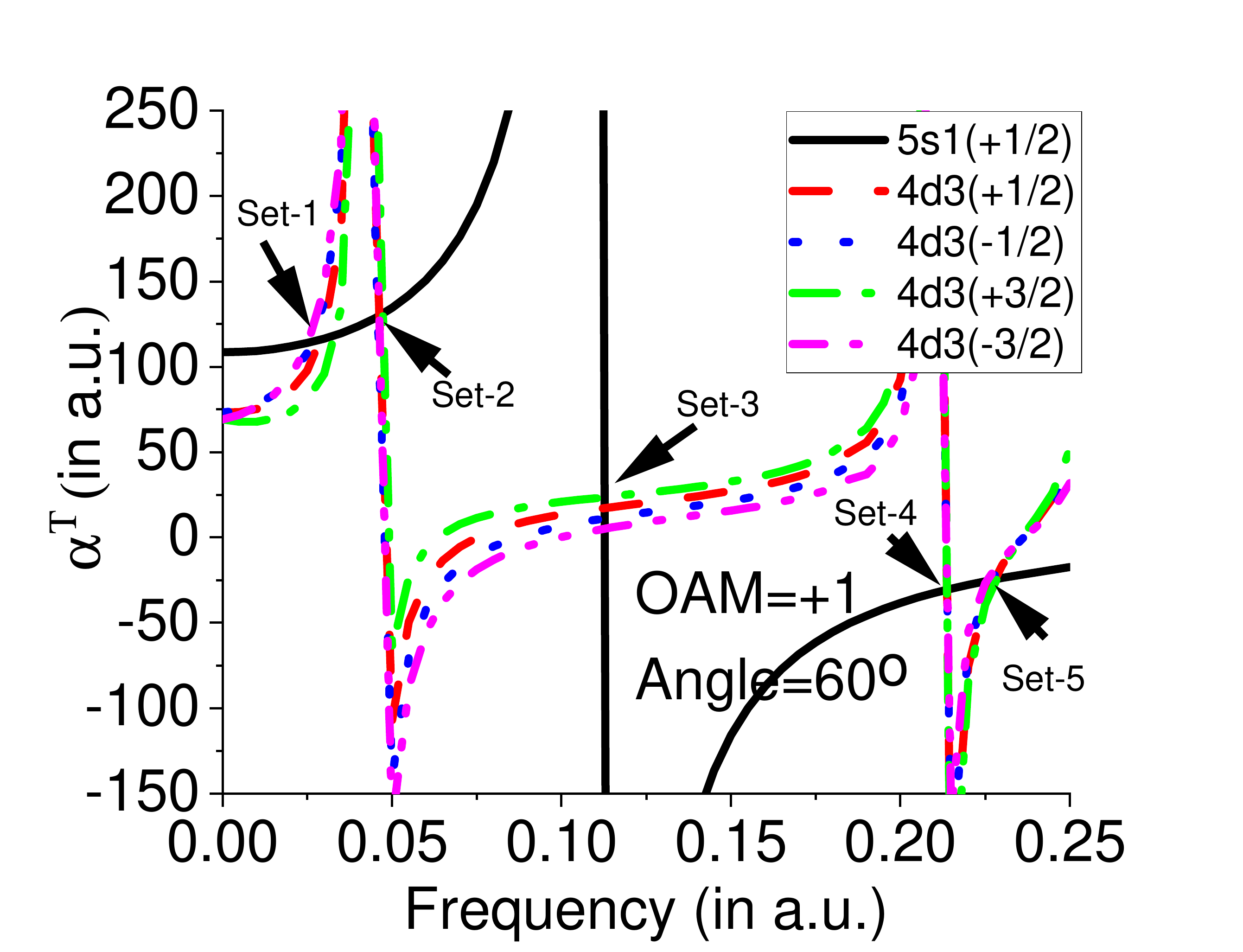}}
{\includegraphics[scale=.30]{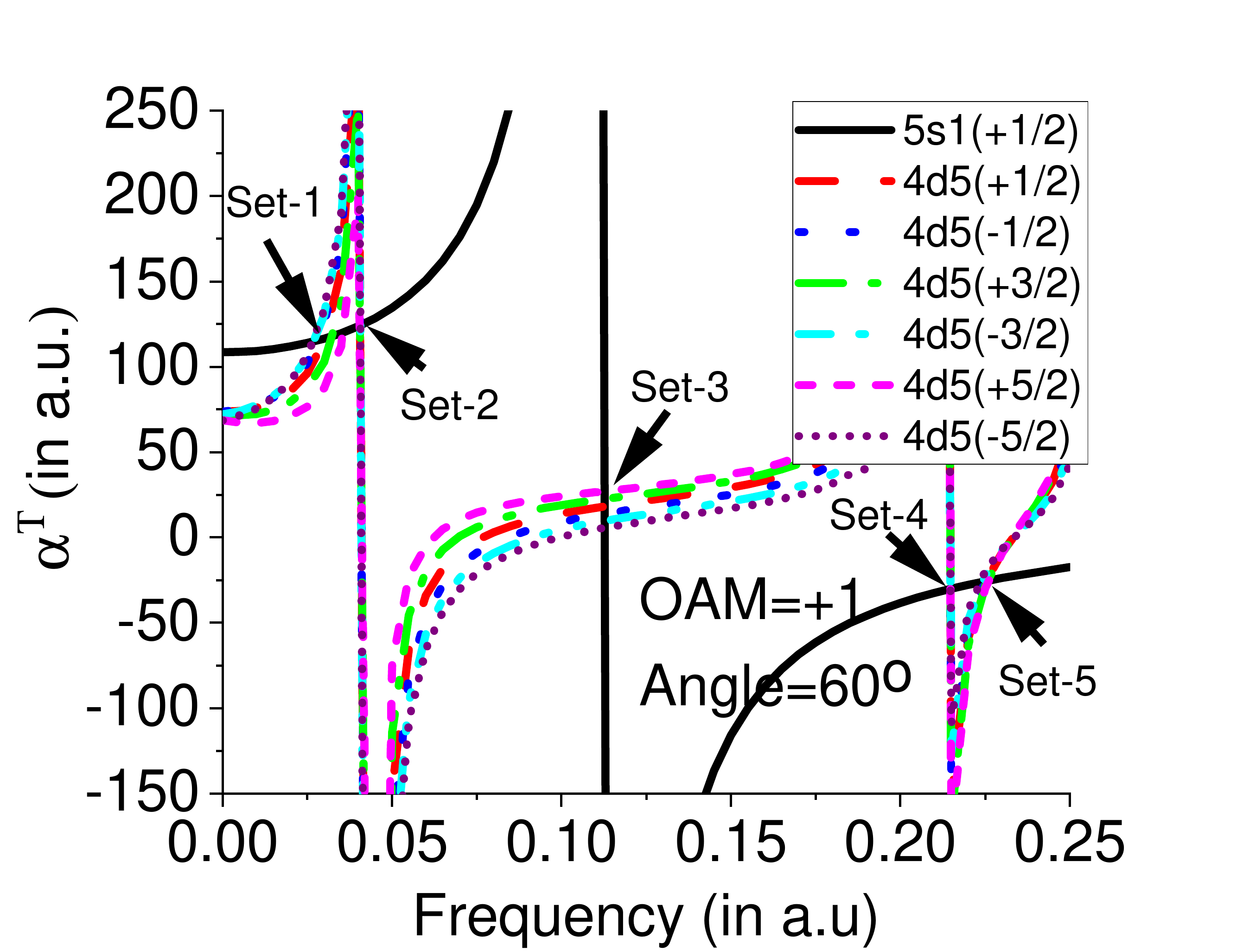}}
	\caption{\label{fig2} Frequency  dependent $\alpha^T$  of  $5S_{\frac{1}{2}}$ and $4D_{\frac{3}{2}, \frac{5}{2}}$ states. LP-FV beam is focused at 60$^\circ$  with OAM=+1.    The values inside the parentheses  indicate  the magnetic components. The corresponding wavelengths at the intersections of the polarizability curves of  $5S_{\frac{1}{2}}$ and $4D_{\frac{3}{2}, \frac{5}{2}}$ states are the magic wavelengths for the clock transitions $5S_{\frac{1}{2}}-4D_{\frac{3}{2}}$ and $5S_{\frac{1}{2}}-4D_{\frac{5}{2}}$. For both these clock transitions, five sets (Set-1 to Set-5) of magic wavelengths are found.}	
\end{figure}

Fig.~\ref{fig2}   displays profiles  of $\alpha^T$  for $5S_{\frac{1}{2}}$ and $4D_{\frac{3}{2}, \frac{5}{2}}$ states of $^{87}$Sr$^+$ at  60$^\circ$  focusing angle for  LP-FV beam  with OAM=+1. However, it should be mentioned that focusing angle does affect  $\alpha^{T}$ values at various frequencies of the beam. Fig.~\ref{fig2} shows a number of intersection points  between $\alpha^T$ profiles of the multiplets of  $5S_{\frac{1}{2}}$ and  $4D_{\frac{3}{2}, \frac{5}{2}}$ states. These intersection points indicate the magic wavelengths at which the differential ac-Stark shift of the  associated clock transition states vanishes.  We observe   five sets (Set-1 to Set-5) of magic wavelengths  for $5S_{\frac{1}{2}}\rightarrow 4D_{\frac{3}{2}, \frac{5}{2}}$ clock transitions within the frequency span (from the near-infrared to the ultra-violet regions) as shown in the figures (see Appendix). This is true for all the magnetic sublevels involed in the transitions.  However, the infrared magic wavelengths  (fall under Set-1 and Set-2) are the most important to trap $^{87}$Sr$^+$ ion due to relatively large $\alpha^T$  values and support the red-detuned trapping scheme useful for frequency standard experiment.  We find that the magic wavelengths for the clock transitions  at Set-1 are maximally affected by the spin-orbit coupling of the beam. For $5S_{\frac{1}{2}}\rightarrow 4D_{\frac{5}{2}}(+3/2)$ transition, the  magic wavelength at Set-1 is 1793.83 nm for either   linearly polarized paraxial LG beam or Gaussian beam.  The results are same as  OAM of the paraxial light does not affect the electronic motion of an cold atom or ion (which is below its recoil limit) at the dipole transition level \cite{Bhowmik2018, Mondal2014}.  However, we observe that for OAM=+1 and +2 of LP-FV beam the magic wavelengths of $5S_{\frac{1}{2}}\rightarrow 4D_{\frac{5}{2}}(+3/2)$ transition at  Set-1 can be varied upto -20\% and -15\%, respectively, compared to the corresponding magic wavelength of Gaussian beam.

\begin{figure}[htb]
{\includegraphics[scale=.30]{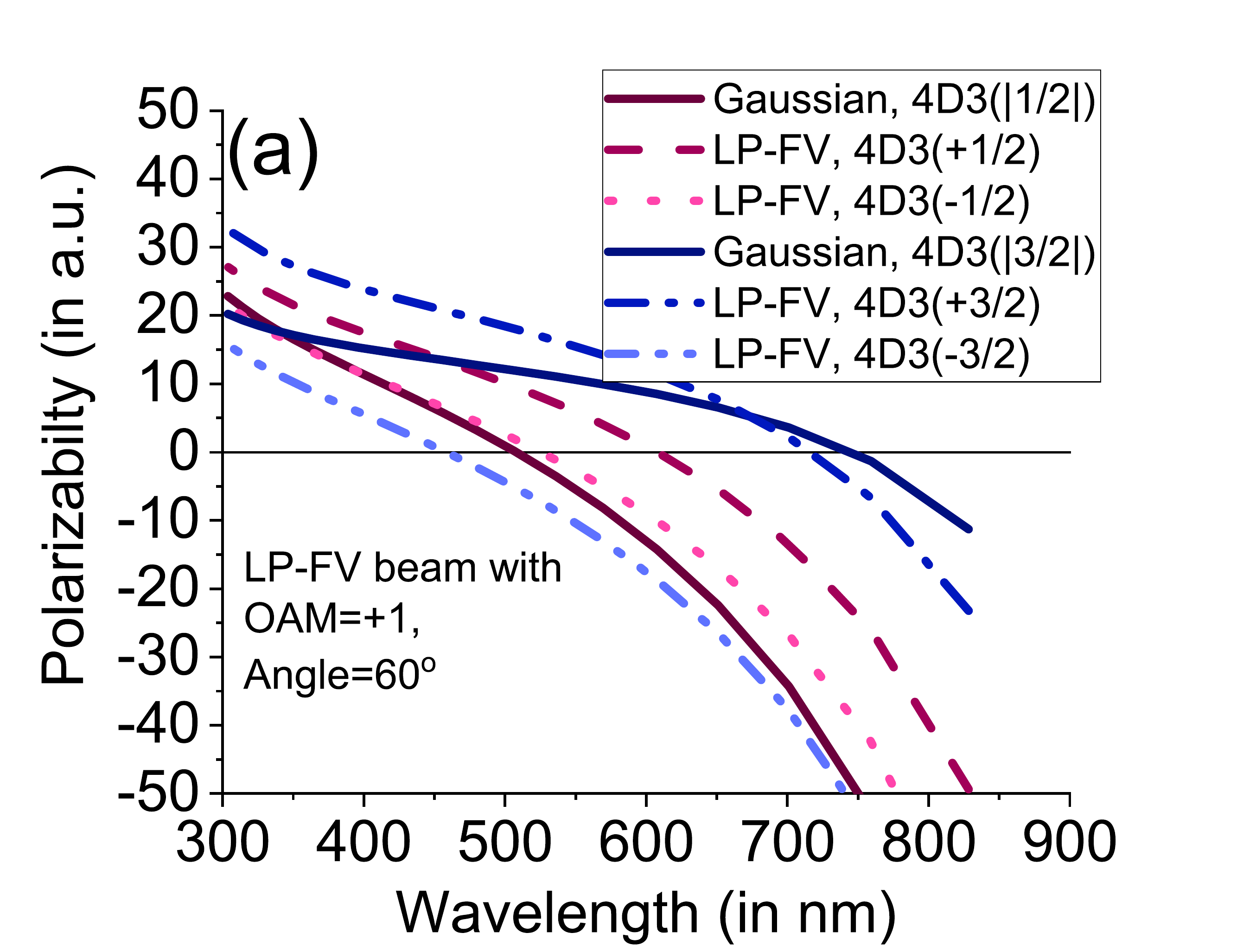}}
{\includegraphics[ scale=.30]{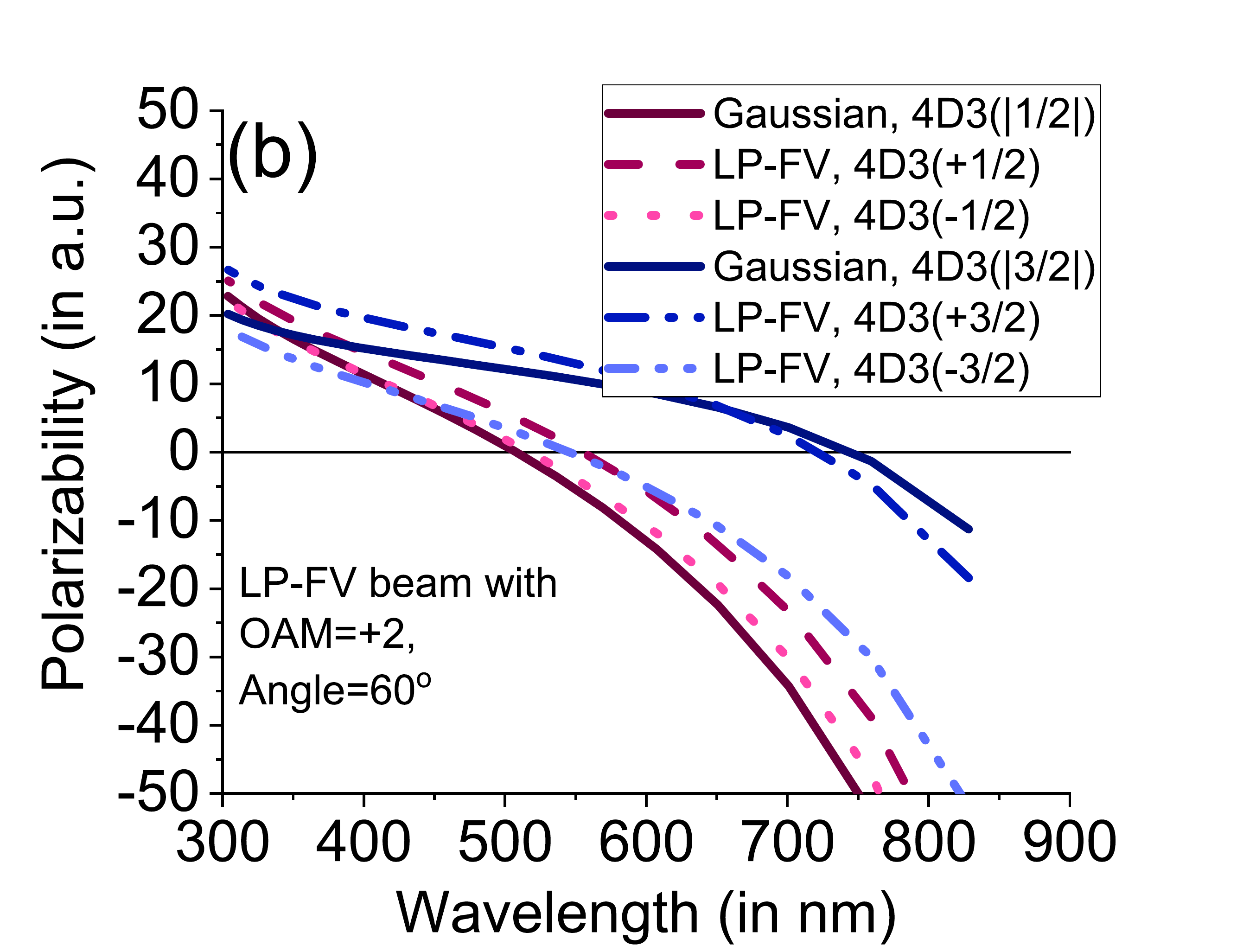}}
\caption{\label{fig3}  Variation of $\alpha^T$ (in a.u.) with wavelength (in nm) for  all the magnetic components of the state   $4D_{\frac{3}{2}}$ for LP-FV beam focused at 60$^\circ$ (with (a) OAM=+1 and (b) OAM=+2) and paraxial Gaussian beam are presented. Wavelengths at which the curves intersects zero-polarizability axis are the tune-out wavelengths.}
\end{figure}
\begin{figure}[htb]
{\includegraphics[ scale=.30]{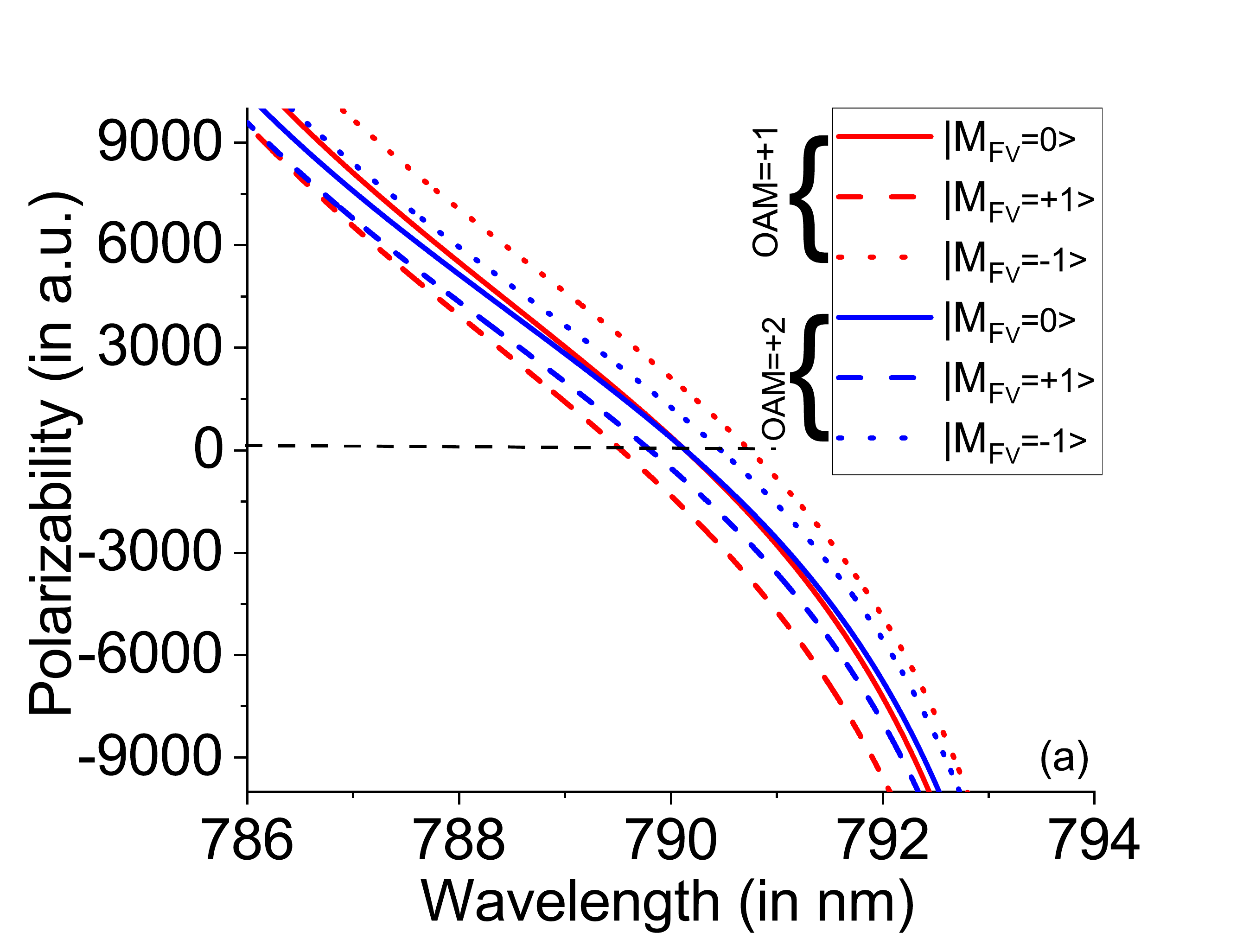}}
{\includegraphics[ scale=.30]{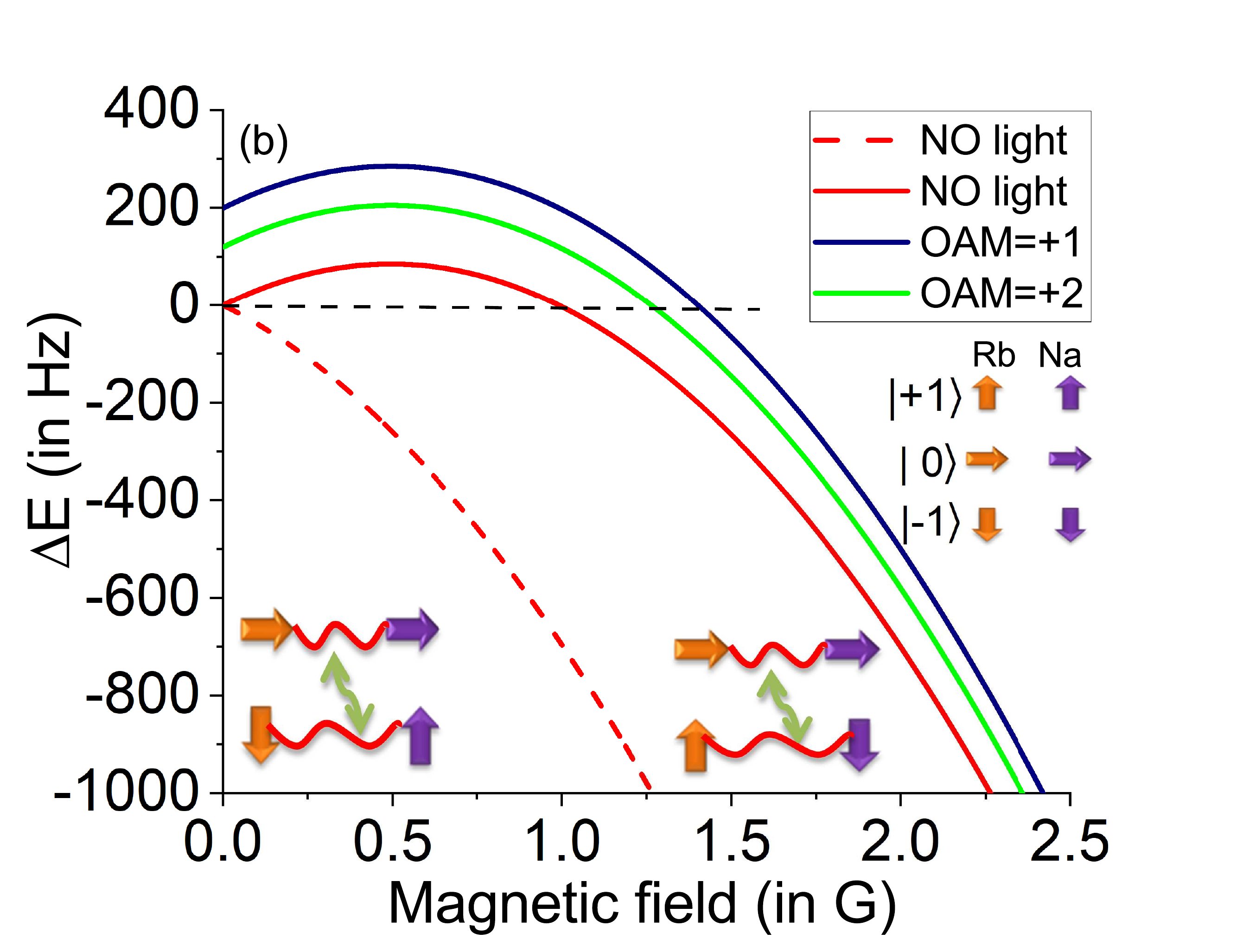}}
\caption{\label{fig4}(a) Variation of $\alpha^T$ (in a.u.) with wavelength (in nm) for the ground state of the $^{87}$Rb  at the hyperfine level $F_V=1$ for the LP-FV beam  focused at 60$^\circ$. (b) Magnetic energy diagram for the two heteronuclear spin-oscillation processes without light shift: $|0,0\rangle \leftrightarrow |-1,1\rangle$ (red dashed) and  $|0,0\rangle \leftrightarrow |1,-1\rangle$ (red solid). Blue solid and green solid curves are magnetic energy variations with OAM=+1 and +2, respectively,  for $|0,0\rangle \leftrightarrow |1,-1\rangle$. The intensity of the LP-FV beam is assumed to be 10 W/m$^2$.}
\end{figure}

Fig.~\ref{fig3} illustrates  wavelength dependence on $\alpha^T$ for the clock transition state $4D_{ \frac{3}{2}}$ of $^{87}$Sr$^+$ due to the external field of  LP-FV beam with OAM=+1 and +2, focused at angle of 60$^\circ$, and linearly polarized paraxial Gaussian beam. As the beam is linearly polarized, the vector component does not arise in $\alpha^T$ for the Gaussian beam. Therefore, the tune-out wavelengths  for  $4D_{ \frac{3}{2}}(M_J=\pm 1/2)$ and $4D_{ \frac{3}{2}}(M_J=\pm 3/2)$ states are degenerated with respect to the sign of $M_J$ and their values are 507.83 nm and 744.12 nm, respectively. These degeneracies are lifted for LP-FV beam. This leads to splitting of the tune-out wavelengths for the oppositely signed $M_J$-levels with respect to the corresponding tune-out wavelengths for the linearly polarized Gaussian beam. The figure shows that non-zero OAM of the LP-FV beam causes red-shifting and blue-shifting of the tune-out wavelengths with respect to the tune-out wavelengths of the Gaussian beam for $M_J=|\frac{1}{2}|$ and $M_J=|\frac{3}{2}|$, respectively. The comparison of panels (a) and (b)  of  Fig.~\ref{fig3} reveals that the separation between the tune-out wavelengths corresponding to $+M_J$ and $-M_J$  of a particular value of $M_J$ decreases with the increasing OAM of the beam. This also indicates the decreasing strength of the spin-orbit coupling with the increasing OAM of the LP-FV beam

We also gauge the effect of the vector POL  arising from  LP-FV beam by considering a    heteronuclear  spin-1 mixture of $^{87}$Rb and $^{23}$Na atoms with spin states $|M_{F_{V1}}\rangle$ and $|M_{F_{V2}}\rangle$, respectively, of  $F_{V1}=F_{V2}=1$ hyperfine levels of their respective ground states $5S_{\frac{1}{2}}$ and $3S_{\frac{1}{2}}$. Here, we  show  how their coherent spin oscillations can be controlled  by the fictitious magnetic field generated from the vector POL due to LP-FV beam.  We calculate the total magnetic energy $E^{|M_{F_{V1}}, M_{F_{V2}}\rangle}(B)$ (from the Breit-Rabi formula  \cite{Vanier1988}) associated with $|M_{F_{V1}}, M_{F_{V2}}\rangle$ state, where $B$ is the externally applied magnetic field.  In this example, we start with   $|0,0\rangle$ mixture of  $^{87}$Rb and $^{23}$Na atoms. Therefore, we  have only two types of spin-changing transition process at  $F_V=1$ level conserving the total magnetization:  $|0,0\rangle \leftrightarrow |-1,1\rangle$ and  $|0,0\rangle \leftrightarrow |1,-1\rangle$. In the presence of a laser beam, the magnitudes of the frequency-dependent Stark-shifts for both the atoms largely depend on the positions of the $D$-line transitions, which are close to 590 nm and 790 nm for the $^{23}$Na and $^{87}$Rb atoms, respectively. Therefore, if a light beam of wavelength nearly 790 nm is used, it  can make the Stark-shift essentially species-selective for $^{87}$Rb, but mostly transparent for $^{23}$Na.  The dynamic polarizability for  the state $|5S_{1/2},F_V=1, M_{F_V}=0,\pm1\rangle$ of $^{87}$Rb is  depicted in Fig~\ref{fig4}(a) for 786-794 nm.  Here a LP-FV beam with intensity $10$ W/cm$^2$  and both OAM=+1, +2 are employed to selectively dresses the energy levels of  $^{87}$Rb near 790.14nm wavelength, which is the tune-out wavelength for $M_{F_V}=0$ state. The figure shows that the Stark shift for $M_{F_V}=\pm 1$ is larger for OAM=+1  compared to OAM=+2 at the wavelength  790.14 nm.

Without the LP-FV beam, the total 
Zeeman  energy difference of the process $|0,0\rangle \leftrightarrow |-1,1\rangle$ is indicated by $\Delta E^{-}(B)=E^{|0,0\rangle}-E^{|-1,1\rangle}$ and of the process $|0,0\rangle \leftrightarrow |1,-1\rangle$ is indicated by $\Delta E^{+}(B)=E^{|0,0\rangle}-E^{|1,-1\rangle}$. These Zeeman energy differences of the heteronuclear spin-oscillations include contributions of both the linear and quadratic Zeeman shifts \cite{Zhang2005}. Since the heteronuclear spin oscillation can only occur near  $\Delta E^{\pm}(B)=0$ \cite{Li2015},  Fig.~\ref{fig4}(b)  suggests that only the oscillation $|0,0\rangle \leftrightarrow |1,-1\rangle$ is possible  at the magnetic field 0.99 G. The other oscillation, $|0,0\rangle \leftrightarrow |-1,1\rangle$, is strongly suppressed at this magnetic field. Now, taking into account  the light shift by the LP-FV beam on $^{87}$Rb atom at the wavelength  790.14 nm, the total Zeeman energy difference of the allowed process is modified to $\Delta E^{+}(B)=(E^{|0,0\rangle}-E^{|1,-1\rangle})+[\delta E_l(M_{F_V}=0)-\delta E_l(M_{F_V}=1)]$, where $\delta E_l$ is the light shift.  In Fig~\ref{fig4}(b), we find that $\Delta E^{+}(B)$ becomes zero at the  magnetic fields 1.40 G and 1.26 G for OAM=+1 and +2, respectively, of LP-FV beam.  Therefore, the external magnetic field that is required for  $|0,0\rangle \leftrightarrow |1,-1\rangle$ transition with $\Delta E^{+}(B)=0$  to take place is changed significantly due to the presence of the LP-FV beam, which accounts the fictitious magnetic field generated by the vector polarizability. This is purely a manifestation of the vector light shift arises from the spin-orbit coupling of the beam.  Also, the spin oscillations process of the heteronuclear  spin-1 mixture of $^{87}$Rb and $^{23}$Na can be controlled externally by changing focusing angle and OAM of LP-FV beam. This process will be very useful for recent experiment using spherical-quadrupole magnetic trap \cite{Fang2019}.
 
 \section{CONCLUSION}
 In summary, we have developed a theoretical formalism to calculate  the dynamic POL  of an  atomic state  in the external field of a LP-FV beam. We have  demonstrated how the OAM and focusing angle of the beam induce the spin-orbit coupling in the beam, which in turn produces vector polarizability in the atomic state even for linearly polarized light.  The fine tuning of magic wavelengths for the clock transitions $5S_{\frac{1}{2}}\rightarrow 4D_{\frac{3}{2}, \frac{5}{2}}$ of  $^{87}$Sr$^{+}$ ion are demonstrated  by controlling  OAM and focusing parameters of LP-FV beam (details are in the Appendix). The vector POL values presented in this paper can be verified experimentally by stimulated Raman spectroscopy \cite{Hu2018} or by measuring the tune-out wavelengths for the magnetic sublevels of same multiplets \cite{Schmidt2016, Trubko2017}. Moreover, we develop a controlling mechanism of heteronuclear spin oscillations of spin-1 mixture of $^{87}$Rb and $^{23}$Na atoms with the fictitious magnetic field generated by LP-FV beam. This mechanism can be used to tune the heteronuclear spin dynamics \cite{Li2020} and   generate entanglement between distinguishable atoms \cite{Shi2006}. We believe that the present theoretical development will give an additional flexibility in the spin-changing interaction process of other multi-species spinor condensates.

\section*{ACKNOWLEDGMENTS} 
A.B. acknowledges Arghya Das for helpful discussions.

\begin{widetext}
\section*{APPENDIX}
In this Appendix, we augment the main text with further details. In section A, we present a brief mathematical description of the scalar, vector, and tensor polarizabilities of a single-valence atomic state at the fine-structure and hyperfine levels. In section B, we show the numerical procedure to calculate the dynamic polarizabilities of $^{87}$Sr$^+$ and $^{87}$Rb. In this Section, we also include the details of the Zeeman energy difference that leads the spin oscillation processes of the spin-1 mixture $^{87}$Rb and $^{23}$Na atoms. In section C, we tabulate a list of magic wavelengths and the corresponding vector and total polarizabilities for the clock transitions $5S_{\frac{1}{2}}\rightarrow 4D_{\frac{3}{2}, \frac{5}{2}}$ of $^{87}$Sr$^+$ ion.

\subsection{Scalar, vector, and tensor  polarizabilities at the  fine-structure and hyperfine levels}

$\alpha_V^0(\omega)  $, $ \alpha_V^1(\omega) $, and $  \alpha_V^2(\omega) $  are the scalar, vector, and tensor components, respectively, of the dynamic valence polarizability $\alpha^V(\omega)$ at the fine-structure level of single-valence state $\Psi_V$. These components are expressed using the sum-over-states approach \cite{Flambaum2008, Arora2007, Lacroute2012, Arora2012, Mitroy2010, Dutta2015} as follows
\setcounter{equation}{0}
\renewcommand{\theequation}{A.\arabic{equation}}
\begin{widetext} 
 \begin{equation}\label{10}
 \alpha_V^0(\omega)=\frac{2}{3(2J_V+1)}\sum_N \frac{|\langle\psi_V||d||\psi_N\rangle|^2\times(\epsilon_N-\epsilon_V)}{(E_N-E_V)^2-\omega^2},
 \end{equation}
\begin{equation}\label{11}
\alpha_V^1(\omega)=-\sqrt{\frac{6J_V}{(J_V+1)(2J_V+1)}}\sum_N (-1)^{J_N+J_V} \left\{\begin{array}{ccc} J_V & 1 & J_V\\ 1 & J_N& 1 \end{array}\right \}\frac{|\langle\psi_V||d||\psi_N\rangle|^2 \times 2\omega}{(E_N-E_V)^2-\omega^2},
\end{equation}

\begin{equation}\label{12}
\alpha_V^2(\omega)=4\sqrt{\frac{5J_V(2J_V-1)}{6(J_V+1)(2J_V+1)(2J_V+3)}}\sum_N (-1)^{J_N+J_V} \left\{\begin{array}{ccc} J_V & 1 & J_N\\ 1 & J_V& 2 \end{array}\right \} \frac{|\langle\psi_V||d||\psi_N\rangle|^2\times(\epsilon_N-\epsilon_V)}{(E_N-E_V)^2-\omega^2}.
\end{equation}
 
\end{widetext}
Here $J_V$ is the total angular momentum and $E_V$ is the ionization potential of $\psi_V$. $\langle\psi_V||d||\psi_N\rangle$ is the reduced  matrix element of the electric dipole operator.

At a hyperfine  level $F_V$ with nuclear spin $I$, the scalar component ($\alpha_{VF}^0(\omega)$) of the valence polarizability is equal to $\alpha_V^0(\omega)$.  This is because of the second-order scalar energy shift does not depend on any hyperfine quantum number \cite{Beloy2009}. But the hyperfine-induced vector ($\alpha_{VF}^{(1)}(\omega)$) and tensor  ($\alpha_{VF}^{(2)}(\omega)$) components  have different angular momentum factors to be multiplied with the above expressions for $\alpha^1_V(\omega)$ and  $\alpha^2_V(\omega)$, respectively \cite{Beloy2009,Dzuba2010,Das2020}:
\begin{equation}\label{13} 
 \alpha_{VF}^{(1)}(\omega)=(-1)^{J_V+F_V+I+1}\left\{\begin{array}{ccc} F_V & J_V & I\\ J_V & F_V& 1 \end{array}\right \}\sqrt{\frac{F_V(2F_V+1)(2J_V+1)(J_V+1)}{J_V(F_V+1)}}\alpha_V^{(1)}(\omega)
 \end{equation}
 and
 \begin{eqnarray}\label{14}
 \alpha_{VF}^{(2)}(\omega)=(-1)^{J_V+F_V+I}\left\{\begin{array}{ccc} F_V & J_V & I\\ J_V & F_V& 2 \end{array}\right \}&& \sqrt{\left(\frac{F_V(2F_V-1)(2F_V+1)}{(2F_V+3)(F_V+1)}\right)}  \nonumber\\ 
&\times &\sqrt{ \left(\frac{(2J_V+3)(2J_V+1)(J_V+1)}{J_V(2J_V-1)}\right)}\alpha_V^{(2)}(\omega).
 \end{eqnarray}
 The total valence polarizability at the hyperfine level ($\alpha_F^V(\omega)$)  takes  the form   as
\begin{equation}\label{15}
\alpha_F^V(\omega) =C_0 \alpha_{VF}^0(\omega)+ C_1\alpha_{VF}^1(\omega)  + C_2\alpha_{VF}^2(\omega),
\end{equation}
where $C_i$s are the coefficients mentioned in the main text.
\subsection{Details of numerical procedure:}
Here, we present the numerical procedure to calculate the dynamic polarizability of an atomic state discussed in the main text.  The electron-correlation significantly affects  the valence polarizability  of an atomic state  due to the loosely bound  valence electron. Therefore, the precise estimations of the scalar, vector, and tensor components of the valence polarizability for an atomic state require correlation-exhaustive many-body treatments with a sophisticated numerical approach.

\subsection*{Calculation of the dynamic polarizability for $^{87}$Sr$^+$ ion}

 To calculate the valence polarizabilities for  $5S_{\frac{1}{2}}$, $4D_{\frac{3}{2}}$, and $4D_{\frac{5}{2}}$ states, we consider  all possible non-negligible dipole matrix elements in   Eq.~(\ref{10}) to ~(\ref{12}).   This sets the highest principle quantum number of  the running index $N$, which is  considered up to 25. According to the significance of each dipole matrix elements to the summations in the equations, many-body calculations of different orders of correlations are performed with negligible compromise in the accuracy of of the polarizability values \cite{Bhowmik2018}. The most dominant  dipole matrix elements are associated with the intermediate states 5$P_{1/2, 3/2}$ to 8$P_{1/2, 3/2}$ and 4$F_{5/2, 7/2}$ to 6$F_{5/2, 7/2}$ as $\psi_N$,  and they are computed using a relativistic coupled-cluster  theory having  cluster operators containing  single, double, and valence triple excitations in linear and non-linear forms \cite{Lindgren1985, Bishop1987, Lindgren1986, Lindgren_1985, Bartlett2007,Dutta2016,  Bhowmik2017,  Das2018, Biswas2018}. Relatively less significant dipole matrix elements are involved with the states 9$P_{1/2, 3/2}$ to 12$P_{1/2, 3/2}$  and 7$F_{5/2, 7/2}$ to 12$F_{5/2, 7/2}$ are evaluated using the 2nd-order relativistic many-body perturbation theory \cite{Lindgren1986}.  The remaining matrix elements contribute  little to the summations of  Eq.~(\ref{10}) to ~(\ref{12}) and are calculated using the Dirac-Fock method. Also,  to achieve a better accuracy in the total dynamic polarizability values, we have utilized the transition energies from the experimental data \cite{NIST}. By employing the above methods, the static scalar polarizabilities $(\alpha_V^0(0))$ of $5S_{\frac{1}{2}}$, $4D_{\frac{3}{2}}$, and  $4D_{\frac{5}{2}}$ states are calculated as 87.68 a.u., 55.92 a.u., and 56.21 a.u., respectively, and static tensor polarizabilities $(\alpha_V^2(0))$ of  $4D_{\frac{3}{2}}$ and  $4D_{\frac{5}{2}}$ states are computed  as $-34.67$ a.u. and $-47.12$ a.u., respectively \cite{Bhowmik2018}.

The core polarizability $\alpha^C (\omega)$ is valence-state independent quantity and it can be calculated quite accurately for a system having noble gas electronic configuration with core polarization corrected dipole matrix elements using  3rd-order relativistic many-body perturbation theory \cite{Johnson1996, Dutta2020}. Our calculation  yields that the static core POL ($\alpha^C(0)$) of the ion is 6.103 a.u.. The  static core-valence parts $\alpha^{VC}(0)$ for the states $5S_{\frac{1}{2}}$, $4D_{\frac{3}{2}}$, and  $4D_{\frac{5}{2}}$  are calculated to be  $-0.25$ a.u., $-0.38$ a.u., and $-0.42$ a.u., respectively. All the magic and tune-out wavelengths calculated for the ion using the above mentioned numerical procedure have maximum uncertainty of around $\pm1\%$. To calculate the uncertainty, we replace the  most important dipole matrix elements of the transitions: $5S_{\frac{1}{2}}\rightarrow 5P_{\frac{1}{2},\frac{3}{2}}$ for $5S_{\frac{1}{2}}$, $4D_{\frac{3}{2}}\rightarrow 5P_{\frac{1}{2},\frac{3}{2}}, 4F_{\frac{5}{2}}$  for $4D_{\frac{3}{2}}$, and $4D_{\frac{5}{2}}\rightarrow 5P_{\frac{3}{2}},4F_{\frac{5}{2}, \frac{7}{2}}$  for $4D_{\frac{5}{2}}$ by the corresponding matrix elements calculated by Safronova \cite{Safronova2010} using single-double all-order method including partial triple excitations and recalculate the magic and tune-out wavelengths. The maximum difference between  these recalculated wavelengths and the corresponding wavelengths obtained from our method gives the uncertainty of our results.

\subsection*{Calculation of the dynamic polarizability for $^{87}$Rb atom}

 To determine the valence polarizability  of the ground state $5S_{\frac{1}{2}}$  of  $^{87}$Rb atom at the  hyperfine level,  Eq.~(\ref{15}) reduces   to  
$\alpha_F^V(\omega) =C_0 \alpha_{VF}^0(\omega)+ C_1\alpha_{VF}^1(\omega)$. The tensor part of the polarizability is zero for $5S_{\frac{1}{2}}$ state. Similar to $^{87}$Sr$^+$ ion, here also, the dipole matrix elements  involved with the states 5$P_{1/2, 3/2}$ to 8$P_{1/2, 3/2}$ and   9$P_{1/2, 3/2}$ to 12$P_{1/2, 3/2}$ are computed using the relativistic coupled-cluster theory and 2nd-order relativistic many-body perturbation theory, respectively.  13$P_{1/2, 3/2}$ to 25$P_{1/2, 3/2}$ states and their associated matrix elements with $5S_{\frac{1}{2}}$ state are calculated using the Dirac-Fock method. To achieve  better accuracy, the energy values of the states, used to calculate the polarizability values, are extracted from the experimental measurements \cite{NIST}. For $5S_{\frac{1}{2}}$ state of $^{87}$Rb atom, we find that the static scalar, core,  and  core-valence polarizability values  are 314.14 a.u.,  9.11 a.u., and   $-0.26$ a.u., respectively. The tune-out wavelengths of the hyperfine spin states of $^{87}$Rb atom, discussed in the main text, have  maximum uncertainty of $\pm 0.1\%$.

\subsection*{Zeeman shift for $^{87}$Rb and $^{23}$Na atoms}
The Zeeman shifts for $^{87}$Rb and $^{23}$Na atoms are evaluated using  $\hat{H}_Z=-\beta \hat{F}_Z+\gamma\hat{F}_Z^2$, where $\beta$ and $\gamma$ are  the  linear  and  quadratic  Zeeman  shifts,  respectively.  $\hat{F}_Z$ represents the $z$-component of the external magnetic field. The coefficients  $\beta$ and $\gamma$ can be obtained  from  the  power  series  expansion  of  the  Breit-Rabi  formula \cite{Vanier1988}. The  fine  structure ($g_J$)  and  nuclear  Land\'e  $g$-factors ($g_I$),  used in the Breit-Rabi  formula, are 2.00233113 and $-0.0009951414$, respectively, for $^{87}$Rb atom, while they are 2.0022960 and $-0.000804610$, respectively, for $^{23}$Na atom \cite{Arimondo1977}. Also, the ground state hyperfine splittings  of $^{87}$Rb and $^{23}$Na atoms, require in the  Breit-Rabi  formula,  have the values 6.8 GHz and 1.7 GHz \cite{Arimondo1977}, respectively.

\subsection{List of magic wavelengths with the corresponding  vector and total polarizabilities}
A list of  magic wavelengths and the corresponding total polarizability values at these wavelengths for the transitions $5S_{\frac{1}{2}}\rightarrow 4D_{\frac{3}{2}, \frac{5}{2}}$ are presented in Table~\ref{I}-Table~\ref{III}. Moreover, at each magic wavelength, values of the vector polarizabilities  of the relevant states  are quoted.  Results are displayed for the OAM=+1 and +2, while the focusing angles of the beam are considered  as  50$^\circ$, 60$^\circ$ and 70$^\circ$. As seen in the tables, all the  transitions  between the magnetic sublevels of $5S_{\frac{1}{2}}$ and  $4D_{\frac{3}{2}, \frac{5}{2}}$ states produce five sets of magic wavelengths except few cases, say, $5S_{\frac{1}{2}}\rightarrow 4D_{\frac{5}{2}}$ transition for OAM=+2.  Depending on the proximity of the resonances, the strength of the vector polarizability of one of the  state dominates over  the other state for a particular value of  magic wavelength. For $5S_{\frac{1}{2}}\rightarrow 4D_{\frac{5}{2}}(+5/2)$ transition, we have found (in table III ) two sets of infrared magic wavelengths are missing at the focusing angle 50$^\circ$ and 60$^\circ$  of the beam, but all five sets of magic wavelengths present at 70$^\circ$,  when the projected beam has OAM=+2. This highlights the direct effect of spin-orbit coupling of linearly polarized vortex  beam on the magic wavelengths. All the tables show that  the magic wavelengths fall in the visible and  ultra-violet region of the electromagnetic spectrum  support the blue-detuned trapping scheme confining the ion in the low intensity region of the  beam  \cite{Kennedy2014, Kuga1997, Wright2000, Arlt2001}.

\begin{table*}[htb]
	\scriptsize
	\caption{Magic wavelengths (in nm) with corresponding polarizabilities (in a.u.) of $^{87}$Sr$^{+}$ for different focusing angles of the linearly polarized vortex  beam (with OAM=+1 and +2) for  transitions $5S_{\frac{1}{2}}(+1/2)  $  $\rightarrow$ $4D_{\frac{3}{2}}(M_J) $. The values in the  parentheses refer vector polarizabilities  at the corresponding magic wavelengths. In the parenthesis (a, b), a and b indicate the vector polarizabilities for $5S_{\frac{1}{2}}(+1/2)$  and $4D_{\frac{3}{2}}(M_J)$ states, respectively, in a.u..}
	\centering
	\begin{tabular}{ccc|cc|cc}
		
		\hline \hline

		State      &$\lambda^{50^\circ}$&$\alpha^{50^\circ}$ & $\lambda^{60^\circ}$&$\alpha^{60^\circ}$&  $\lambda^{70^\circ}$&$\alpha^{70^\circ}$
		\\ [0.2ex]
		\hline 
		($4D_{\frac{3}{2}}(M_J)$)& \multicolumn{6}{c}{\textbf{OAM=+1}} \\  
		\hline  
	$(+\frac{1}{2})$	&	1627.26	&	113.53	(-0.16, -7.20)	&	1598.71	&	115.45	(-0.16, -7.81)	&	1571.15	&	117.37	(-0.16, -7.27)	\\
	&	992.67	&	126.99	(-0.33, 12.91)	&	981.97	&	129.94	(-0.34, 13.95)	&	981.97	&	132.77	(-0.32, 13.14)	\\
	&	403.57	&	17.32	(158.00, 3.10)	&	403.57	&	17.32	(156.69, 3.07)	&	403.57	&	17.32	(158.72, 2.91)	\\
	&	213.21	&	-29.88	(-0.13, -0.93)	&	213.21	&	-30.79	(-0.13, -0.84)	&	213.21	&	-31.81	(-0.13, -0.24)	\\
	&	200.01	&	-24.11	(-0.10, -1.37)	&	200.45	&	-25.01	(-0.10, -1.43)	&	200.45	&	-26.03	(-0.09, -1.24)	\\
$(-\frac{1}{2})$	&	1786.80	&	111.60	(-0.15, 5.20)	&	1752.44	&	114.54	(-0.15, 6.21)	&	1752.44	&	116.47	(-0.14, 5.52)	\\
	&	1003.60	&	126.99	(-0.33, -12.51)	&	992.67	&	129.94	(-0.34, -13.25)	&	992.67	&	131.86	(-0.32, -12.93)	\\
	&	403.57	&	10.53	(158.00, -3.10)	&	403.57	&	10.53	(156.69, -3.07)	&	403.57	&	11.54	(158.72, -2.91)	\\
	&	213.21	&	-29.88	(-0.13, 0.93)	&	213.21	&	-30.79	(-0.13, 0.84)	&	213.21	&	-31.81	(-0.13, 0.24)	\\
	&	200.45	&	-25.01	(-0.10, 1.47)	&	200.90	&	-25.01	(-0.10, 1.43)	&	201.34	&	-26.03	(-0.09, 1.64)	\\
$(+\frac{3}{2})$	&	1368.27	&	116.47	(-0.19, -40.15)	&	1389.13	&	118.39	(-0.20, -37.93)	&	1406.28	&	120.32	(-0.19, -34.55)	\\
	&	961.25	&	128.92	(-0.35, 43.73)	&	961.25	&	130.84	(-0.37, 43.90)	&	961.25	&	133.79	(-0.34, 42.50)	\\
	&	403.57	&	23.09	(158.00, 9.25)	&	403.57	&	23.09	(156.69, 9.26)	&	403.57	&	23.09	(158.72, 8.75)	\\
	&	213.21	&	-29.88	(-0.13, -2.78)	&	212.71	&	-30.79	(-0.13, -18.52)	&	213.21	&	-31.81	(-0.13, -2.64)	\\
	&	199.58	&	-24.11	(-0.10, -4.33)	&	199.58	&	-25.01	(-0.10, -5.23)	&	199.58	&	-25.01	(-0.08, -3.46)	\\
$(-\frac{3}{2})$	&	1815.27	&	111.60	(-0.15, 15.61)	&	1815.27	&	113.53	(-0.21, 15.65)	&	1815.27	&	116.47	(-0.13, 14.78)	\\
	&	981.97	&	128.01	(-0.34, -41.81)	&	992.67	&	128.92	(-0.34, -40.55)	&	992.67	&	131.86	(-0.32, -38.00)	\\
	&	403.57	&	4.75	(158.00, -9.25)	&	403.57	&	4.75	(156.69, -9.26)	&	403.57	&	5.77	(158.72, -8.75)	\\
	&	213.21	&	-29.88	(-0.13, 2.78)	&	213.21	&	-30.79	(-0.13, 2.80)	&	213.21	&	-31.81	(-0.13,  2.64)	\\
	&	202.14	&	-25.01	(-0.10, 6.21)	&	202.14	&	-26.03	(-0.10, 6.13)	&	201.79	&	-26.03	(-0.09, 5.29)	\\

		\hline 
		($4D_{\frac{3}{2}}(M_J)$) & \multicolumn{6}{c}{\textbf{OAM=+2}} \\  
		\hline
		$(+\frac{1}{2})$	&	2312.86	&	103.00	(-0.05, -1.66)	&	2201.13	&	104.92	(-0.05, -1.91)	&	2052.40	&	107.75	(-0.06, -2.15)	\\
	&	1047.43	&	117.37	(-0.15, -5.16)	&	1035.53	&	119.30	(-0.16, -1.46)	&	1035.53	&	122.24	(-0.16, -1.43)	\\
	&	405.73	&	13.47	(115.64, 1.51)	&	403.57	&	14.37	(86.72, 1.59)	&	405.73	&	14.37	(119.91, 1.56)	\\
	&	213.21	&	-27.96	(-0.06, -0.55)	&	213.21	&	-28.86	(-0.07, -0.48)	&	213.21	&	-28.86	(-0.06, -0.47)	\\
	&	199.66	&	-22.18	(-0.05, -0.71)	&	200.01	&	-23.09	(-0.05, -0.74)	&	199.58	&	-23.09	(-0.05, -0.61)	\\
$(-\frac{1}{2})$	&	2559.74	&	101.98	(-0.04, 1.43)	&	2360.80	&	104.92	(-0.05, 1.71)	&	2201.13	&	107.75	(-0.06, 1.88)	\\
	&	1035.53	&	117.37	(-0.16, 1.38)	&	1035.53	&	119.30	(-0.16, 1.46)	&	1035.53	&	122.24	(-0.16, 1.43)	\\
	&	403.57	&	10.53	(82.11, -1.51)	&	403.57	&	10.53	(86.72, -1.59)	&	405.73	&	11.54	(119.91, -1.56)	\\
	&	213.21	&	-27.96	(-0.06, 0.55)	&	213.21	&	-28.86	(-0.07, 0.48)	&	213.21	&	-28.86	(-0.06, 0.47)	\\
	&	199.58	&	-22.18	(-0.05, 0.39)	&	200.01	&	-23.10	(-0.05, 0.74)	&	200.01	&	-23.09	(-0.05, 0.69)	\\
$(+\frac{3}{2})$	&	1276.28	&	111.60	(-0.11, -33.13)	&	1276.28	&	112.62	(-0.11, -35.00)	&	1290.75	&	115.45	(-0.11, -28.42)	\\
	&	943.34	&	122.24	(-0.18, 22.70)	&	943.34	&	124.17	(-0.20, 23.98)	&	943.34	&	126.99	(-0.19, 23.54)	\\
	&	405.73	&	19.24	(115.64, 4.52)	&	405.73	&	19.24	(122.13, 4.79)	&	405.73	&	19.24	(119.91, 4.70)	\\
	&	213.21	&	-27.96	(-0.06, -1.37)	&	213.21	&	-28.86	(-0.07, -1.45)	&	213.21	&	-28.86	(-0.06, -1.42)	\\
	&	200.90	&	-22.18	(-0.05, -2.36)	&	200.90	&	-23.09	(-0.05, -2.50)	&	200.45	&	-23.09	(-0.05, -2.25)	\\
$(-\frac{3}{2})$	&	1428.32	&	108.77	(-0.09, 16.18)	&	1451.06	&	109.68	(-0.09, 16.10)	&	1474.54	&	112.62	(-0.09, 14.67)	\\
	&	951.22	&	121.22	(-0.18, -22.19)	&	951.22	&	123.15	(-0.19, -23.54)	&	951.22	&	126.99	(-0.19, -23.01)	\\
	&	403.57	&	9.62	(82.11, -4.51)	&	403.57	&	9.62	(86.72, -4.75)	&	405.73	&	9.62	(119.91, -4.70)	\\
	&	213.21	&	-27.96	(-0.06, 1.37)	&	213.21	&	-28.86	(-0.07, 1.45)	&	213.21	&	-28.86	(-0.06, 1.42)	\\
	&	202.14	&	-23.09	(-0.05, 2.90)	&	202.14	&	-24.11	(-0.05, 3.06)	&	201.79	&	-24.11	(-0.05, 2.84)	\\

		\hline
		\hline
		\label{table:nonlin} 
		\label{I}

	\end{tabular}
\end{table*}
\begin{table*}[htb]
	\scriptsize
	\caption{Magic wavelengths (in nm) with corresponding polarizabilities (in a.u.) of $^{87}$Sr$^{+}$ for different focusing angles of linearly polarized vortex  beam with OAM=+1 for  transitions $5S_{\frac{1}{2}}(+1/2)  $  $\rightarrow$ $4D_{\frac{5}{2}}(M_J) $. The values in the  parentheses refer vector polarizabilities at the corresponding magic wavelengths. In the parenthesis (a, b), a and b indicate the vector polarizabilities for $5S_{\frac{1}{2}}(+1/2)$ and $4D_{\frac{5}{2}}(M_J)$ states, respectively, in a.u..}
	\centering
	\begin{tabular}{ccc|cc|cc}
		
		\hline \hline

		State      &$\lambda^{50^\circ}$&$\alpha^{50^\circ}$ & $\lambda^{60^\circ}$&$\alpha^{60^\circ}$&  $\lambda^{70^\circ}$&$\alpha^{70^\circ}$
		\\ [0.2ex]
		\hline 
		($4D_{\frac{5}{2}}(M_J)$)& \multicolumn{6}{c}{\textbf{OAM=+1}}\\  
		\hline  
		$(+\frac{1}{2})$	&	1544.52	&	114.54	(-0.17, -4.39)	&	1523.86	&	116.47	(-0.18, -4.56)	&	1523.86	&	118.39	(-0.17, -4.31)	\\
	&	1122.25	&	122.24	(-0.28, -5.30)	&	1122.25	&	124.17	(-0.28, -5.33)	&	1122.25	&	126.99	(-0.26, -5.46)	\\
	&	403.57	&	18.22	(158.00, 2.14)	&	403.57	&	18.22	(156.69, 2.15)	&	403.57	&	18.22	(158.72, 2.03)	\\
	&	212.22	&	-29.88	(-0.13, 1.44)	&	212.22	&	-29.88	(-0.13, 1.45)	&	212.22	&	-30.79	(-0.12, 1.36)	\\
	&	202.14	&	-25.01	(-0.10, -0.89)	&	202.59	&	-26.03	(-0.10, -1.03)	&	202.14	&	-26.03	(-0.10, -0.91)	\\
$(-\frac{1}{2})$	&	1656.85	&	113.53	(-0.16, 3.77)	&	1656.85	&	114.54	(-0.16, 3.78)	&	1627.26	&	117.37	(-0.15, 3.72)	\\
	&	1122.25	&	122.24	(-0.28, 5.30)	&	1122.25	&	124.17	(-0.28, 5.33)	&	1122.25	&	126.99	(-0.26, 5.46)	\\
	&	403.57	&	13.47	(158.00, -2.14)	&	403.57	&	13.47	(156.69, -2.15)	&	403.57	&	14.37	(158.72, -2.03)	\\
	&	212.22	&	-29.88	(-0.13, -1.44)	&	212.22	&	-30.79	(-0.13, -1.45)	&	212.22	&	-30.79	(-0.12, -1.36)	\\
	&	202.59	&	-25.01	(-0.10, 1.02)	&	202.59	&	-26.03	(-0.10, 1.03)	&	202.59	&	-26.94	(-0.10, 0.96)	\\
$(+\frac{3}{2})$	&	1406.28	&	115.45	(-0.20, -18.64)	&	1428.32	&	117.37	(-0.20, -17.34)	&	1428.32	&	120.32	(-0.19, -16.37)	\\
	&	1122.25	&	122.24	(-0.28 -15.94)	&	1122.25	&	124.17	(-0.28, -15.99)	&	1122.25	&	126.99	(-0.26, -15.09)	\\
	&	403.57	&	23.09	(158.00, 6.52)	&	403.57	&	22.07	(156.69, 6.44)	&	403.57	&	22.07	(158.72, 6.10)	\\
	&	212.22	&	-28.86	(-0.13, 4.33)	&	212.22	&	-30.79	(-0.13, 4.34)	&	212.22	&	-30.79	(-0.12, 4.10)	\\
	&	201.79	&	-25.01	(-0.10, -2.68)	&	201.79	&	-25.01	(-0.10, -2.69)	&	201.79	&	-26.03	(-0.10, -2.56)	\\
$(-\frac{3}{2})$	&	1712.91	&	112.62	(-0.15, 10.46)	&	1687.53	&	114.54	(-0.16, 10.86)	&	1687.53	&	117.37	(-0.15, 10.26)	\\
	&	1122.25	&	122.24	(-0.28, 15.94)	&	1122.25	&	124.17	(-0.28, 15.99)	&	1122.25	&	126.99	(-0.26, 15.09)	\\
	&	403.57	&	9.62	(158.00, -6.52)	&	403.57	&	9.62	(156.69, -6.44)	&	403.57	&	9.62	(158.72, -6.10)	\\
	&	212.22	&	-29.88	(-0.13, -4.33)	&	212.22	&	-30.79	(-0.13, -4.34)	&	212.22	&	-30.79	(-0.12, -4.10)	\\
	&	203.05	&	-26.03	(-0.10, 3.59)	&	203.50	&	-26.94	(-0.10, 4.04)	&	203.50	&	-26.94	(-0.10, 3.82)	\\
$(+\frac{5}{2})$	&	1276.28	&	119.30	(-0.22, -46.32)	&	1276.28	&	120.32	(-0.22, -46.44)	&	1309.29	&	123.15	(-0.21, -36.17)	\\
	&	1122.25	&	122.24	(-0.28, -26.57)	&	1122.25	&	124.17	(-0.28, -26.64)	&	1122.25	&	126.99	(-0.26, -25.16)	\\
	&	403.57	&	26.93	(158.00, 10.70)	&	403.57	&	26.94	(156.69, 10.73 )	&	403.57	&	26.94	(158.72, 10.15)	\\
	&	212.22	&	-28.86	(-0.13, 7.23)	&	212.22	&	-30.79	(-0.13, 7.24)	&	212.22	&	-30.79	(-0.12, 6.84)	\\
	&	201.79	&	-25.01	(-0.10, -4.48)	&	201.79	&	-26.03	(-0.10, -4.49)	&	201.79	&	-26.03	(-0.10, -4.24)	\\
$(-\frac{5}{2})$	&	1687.53	&	112.62	(-0.16, 18.04)	&	1712.91	&	114.54	(-0.15, 17.48)	&	1712.91	&	117.37	(-0.14, 16.51)	\\
	&	1122.25	&	122.24	(-0.28, 26.57)	&	1122.25	&	124.17	(-0.28, 26.64)	&	1122.25	&	126.99	(-0.26, 25.16)	\\
	&	403.57	&	5.77	(158.00, -10.70)	&	403.57	&	5.77	(156.69, -10.73)	&	403.57	&	6.68	(158.72, -10.15)	\\
	&	212.22	&	-29.88	(-0.13, -7.23)	&	212.22	&	-29.88	(-0.13, -7.24)	&	212.22	&	-30.79	(-0.12, -6.84)	\\
	&	204.41	&	-26.03	(-0.10, 8.15)	&	203.95	&	-26.03	(-0.10, 7.46)	&	203.95	&	-26.93	(-0.10, 7.03)	\\

		\hline
		\hline
		\label{table:nonlin} 
		\label{II}

	\end{tabular}
\end{table*}

\begin{table*}[htb]
	\scriptsize
	\caption{Magic wavelengths (in nm) with corresponding polarizabilities (in a.u.) of $^{87}$Sr$^{+}$ for different focusing angles of the linearly polarized vortex  beam with OAM=+2 for  transitions $5S_{\frac{1}{2}}(+1/2)  $  $\rightarrow$ $4D_{\frac{5}{2}}(M_J) $. The values in the  parentheses refer vector polarizabilities  at the corresponding magic wavelengths. In the parenthesis (a, b), a and b indicate the vector polarizabilities for $5S_{\frac{1}{2}}(+1/2)$  and $4D_{\frac{5}{2}}(M_J)$ states, respectively, in a.u..}
	\centering
	\begin{tabular}{ccc|cc|cc}
		
		\hline \hline

		State      &$\lambda^{50^\circ}$&$\alpha^{50^\circ}$ & $\lambda^{60^\circ}$&$\alpha^{60^\circ}$&  $\lambda^{70^\circ}$&$\alpha^{70^\circ}$
		\\ [0.2ex]
		\hline 
		($4D_{\frac{5}{2}}(M_J)$)& \multicolumn{6}{c}{\textbf{OAM=+2}}\\  
		\hline  
		
		$(+\frac{1}{2})$	&	2360.80	&	103.00	(-0.05, -0.93)	&	2149.21	&	105.83	(-0.05, -1.17)	&	2052.40	&	107.75	(-0.06, -1.25)	\\
	&	1122.25	&	114.54	(-0.14, -2.61)	&	1122.25	&	116.47	(-0.14, -2.75)	&	1122.25	&	119.30	(-0.14, -2.70)	\\
	&	405.73	&	14.37	(115.64, 1.05)	&	405.73	&	15.39	(122.13, 1.11)	&	405.73	&	15.39	(119.91, 1.09)	\\
	&	212.22	&	-26.94	(-0.06, 0.71)	&	212.22	&	-27.96	(-0.07, 0.73)	&	212.22	&	-28.86	(-0.06, 0.73)	\\
	&	201.34	&	-23.09	(-0.05,  -0.40)	&	201.79	&	-23.09	(-0.05, -0.44)	&	201.34	&	-24.11	(-0.05, -0.42)	\\
$(-\frac{1}{2})$	&	2489.80	&	103.00	(-0.05, 0.87)	&	2278.17	&	104.81	(-0.05, 1.05)	&	2149.21	&	107.75	(-0.06, 1.15)	\\
	&	1122.25	&	114.54	(-0.14, 2.61)	&	1122.25	&	116.47	(-0.14, 2.75)	&	1122.25	&	119.30	(-0.14, 2.70)	\\
	&	405.73	&	12.45	(115.64, -1.05)	&	403.57	&	13.24	(86.72, -1.11)	&	405.73	&	13.47	(119.91, -1.09)	\\
	&	212.22	&	-26.94	(-0.06, -0.71)	&	212.02	&	-27.39	(-0.07,  2.09)	&	212.22	&	-28.86	(-0.06, -0.73)	\\
	&	201.79	&	-23.09	(-0.05, 0.44)	&	201.88	&	-23.66	(-0.05, 0.44)	&	201.79	&	-24.11	(-0.05,  0.46)	\\
$(+\frac{3}{2})$	&	1544.52	&	107.75	(-0.09, -6.45)	&	1544.52	&	108.77	(-0.09, -6.82)	&	1523.86	&	111.60	(-0.09, -6.93)	\\
	&	1122.25	&	114.54	(-0.14, -7.81)	&	1122.25	&	116.47	(-0.14, -8.25)	&	1122.25	&	119.30	(-0.14, -8.10)	\\
	&	405.73	&	18.22	(115.64, 3.15)	&	405.73	&	18.22	(122.13, 3.34)	&	405.73	&	18.22	(119.91, 3.27)	\\
	&	212.22	&	-26.94	(-0.06, 2.13)	&	212.22	&	-27.96	(-0.07, 2.25)	&	212.22	&	-28.86	(-0.06, 2.20)	\\
	&	201.79	&	-23.09	(-0.05, -1.31)	&	201.79	&	-23.09	(-0.05, -1.38)	&	201.79	&	-24.11	(-0.05, -1.37)	\\
$(-\frac{3}{2})$	&	1752.44	&	105.83	(-0.07, 4.85)	&	1712.91	&	107.75	(-0.07, 5.41)	&	1712.91	&	109.68	(-0.07, 5.31)	\\
	&	1122.25	&	114.54	(-0.14, 7.81)	&	1122.25	&	116.47	(-0.14, 8.25)	&	1122.25	&	119.30	(-0.14, 8.10)	\\
	&	405.73	&	11.54	(115.64, -3.15)	&	403.57	&	11.54	(86.72, -3.34)	&	405.73	&	11.54	(119.91, -3.27)	\\
	&	212.22	&	-26.94	(-0.06, -2.13)	&	212.22	&	-27.96	(-0.07, -2.25)	&	212.22	&	-28.86	(-0.06, -2.20)	\\
	&	202.59	&	-23.09	(-0.05, 1.50)	&	202.59	&	-24.11	(-0.05, 1.59)	&	202.59	&	-24.11	(-0.05, 1.56)	\\
$(+\frac{5}{2})$	&		&			&		&			&	1192.76	&	117.37	(-0.13, -37.85)	\\
	&		&			&		&			&	1136.24	&	119.29	(-0.14, -41.59)	\\
	&	405.73	&	22.07	(115.64, 5.25)	&	405.73	&	22.07	(122.13, 5.55)	&	405.73	&	22.07	(119.91, 5.45)	\\
	&	212.22	&	-26.94	(-0.06, 3.54)	&	212.22	&	-27.96	(-0.07, 3.73)	&	212.22	&	-28.86	(-0.06, 3.67)	\\
	&	203.05	&	-23.09	(-0.05, -2.72)	&	202.59	&	-24.11	(-0.05, -2.65)	&	202.59	&	-24.11	(-0.05, -2.60)	\\
$(-\frac{5}{2})$	&	1276.28	&	110.70	(-0.11, 22.70)	&	1290.75	&	112.62	(-0.11, 21.72)	&	1328.38	&	114.54	(-0.11, 18.54)	\\
	&	1122.25	&	114.54	(-0.14, 13.02)	&	1122.25	&	116.47	(-0.14, 13.74)	&	1122.25	&	119.30	(-0.14, 13.50)	\\
	&	405.73	&	11.54	(115.64, -5.25)	&	403.57	&	11.54	(86.72, -5.55)	&	405.73	&	11.54	(119.91, -5.45)	\\
	&	212.22	&	-26.94	(-0.06, -3.54)	&	212.22	&	-27.96	(-0.07, -3.73)	&	212.22	&	-28.86	(-0.06, -3.67)	\\
	&	204.78	&	-24.11	(-0.05, 4.29)	&	204.41	&	-24.11	(-0.05, 4.23)	&	203.95	&	-25.01	(-0.05, 3.77)	\\

		\hline
		\hline
		\label{table:nonlin} 
		\label{III}

	\end{tabular}
\end{table*}
\end{widetext}

\end{document}